\documentclass[12pt,preprint]{aastex}

\newcounter{address}
\newcommand{\latin}[1]{{#1}}
\newcommand{\ie}{\latin{i.e.}}
\newcommand{\eg}{\latin{e.g.}}
\newcommand{\cf}{\latin{c.f.}}

\newcommand{\vs}{\latin{vs.}}
\newcommand{\Halpha}{\ensuremath{\mathrm{H}\alpha}}
\newcommand{\Dcl} {\ensuremath{D_\mathrm{cl}}}
\newcommand{\Rvir} {\ensuremath{R_\mathrm{vir}}}
\newcommand{\N}{\ensuremath{N_\mathrm{gal}}}

\begin{document}
\title{ What galaxies know about their nearest cluster }
\author{
  Alejandro~D.~Quintero\altaffilmark{\ref{NYU}},
  Andreas~A.~Berlind\altaffilmark{\ref{NYU}},
  Michael~R.~Blanton\altaffilmark{\ref{NYU}},
  David~W.~Hogg\altaffilmark{\ref{NYU},\ref{email}} }

\setcounter{address}{1}
\altaffiltext{\theaddress}{\stepcounter{address}\label{NYU}
Center for Cosmology and Particle Physics, Department of Physics, New
York University, 4 Washington Place, New York, NY 10003}
\altaffiltext{\theaddress}{\stepcounter{address}\label{email}
To whom correspondence should be addressed: \texttt{david.hogg@nyu.edu}}

\begin{abstract}
We investigate the extent to which galaxies' star-formation histories
and morphologies are determined by their clustocentric distance and
their nearest cluster's richness.  We define clustocentric distance as
the transverse projected distance between a galaxy and its nearest
cluster center within $\pm 1000~\mathrm{km\,s^{-1}}$.  We consider a
cluster to be a bound group of at least $\N$ galaxies brighter than
$M_r=-19$~mag; we employ different richness criteria of $\N=5$, 10,
and 20.  We look at three tracers of star-formation history ($^{0.1}i$
band absolute magnitude $M_{^{0.1}i}$, $^{0.1}[g-r]$ color, and
\Halpha\ emission line equivalent width) and two indicators of galaxy
morphology (surface brightness and radial concentration) for 52,569
galaxies in the redshift range of $0.015<z<0.068$.  We find that our
morphology indicators (surface brightness and concentration) relate to
the clustocentric distance only indirectly through their relationships
with stellar population and star formation rate.  Galaxies that are
near the cluster center tend to be more luminous, redder and have
lower \Halpha\ EW (\ie, lower star-formation rates) than those that
lie near or outside the virial radius of the cluster.  The detailed
relationships between these galaxy properties and clustocentric
distance depend on cluster richness.  For richer clusters, we find
that \textsl{(i)}~the transition in color and \Halpha\ EW from cluster
center to field values is more abrupt and occurs closer to the cluster
virial radius, and \textsl{(ii)}~the color and \Halpha\ EW
distributions are overall narrower than in less rich clusters.  We
also find that the radial gradient seen in the luminosity distribution
is strongest around the smaller clusters and decreases as the cluster
richness (and mass) increases.  We find there is a ``characteristic
distance'' at around one virial radius (the infall region) where the
change with radius of galaxy property distributions is most dramatic,
but we find no evidence for infall-triggered star bursts.  These
results suggest that galaxies ``know'' the distance to, and the size
of, their nearest cluster and they express this information in their
star-formation histories.
\end{abstract}


\section{Introduction}

The statistical properties of galaxies are closely related to the
densities of their surrounding environments.  Since regions of the
Universe with different densities evolve at different rates, we expect
these environment dependencies to contain crucial information about
galaxy formation and evolution.

Much of previous environment-related work focused on the relationship
between morphology and environment (\citealt{hubble36a, oemler74a,
dressler80a, postman84a}; or the more recent work of
\citealt{hermit96a, guzzo97a, giuricin01a, trujillo02a}).  All these
works find that bulge-dominated galaxies are more strongly clustered
than disk-dominated galaxies.  Spectroscopic and photometric
properties of galaxies are strongly correlated with morphology so it
is not surprising to find that the spectroscopic and photometric
properties of galaxies are also functions of clustering
\citep{kennicutt83a, hashimoto98a, balogh01a, martinez02a, lewis02a,
norberg02a, hogg03b, hogg04a, kauffmann04a, zehavi05b}.  While trying
to understand \emph{why} these properties relate with environment, it
is important to ask which properties are correlated with environment
independently of the others.  Previous work has found that, out of
color, luminosity, surface brightness, and S\'ersic index
(concentration), the color and luminosity of a galaxy appear to be the
only properties that are directly related to the local overdensity
\citep{blanton05b}.  Surface brightness and concentration appear to be
related to the environment only through their relationships with color
and luminosity.

The clustocentric distance---the distance to the nearest rich cluster
center---is a fundamental and precisely measurable environment
indicator; indeed it can be measured at much higher signal-to-noise
than estimates of overdensity on fixed scales.  Previous
observational results that use the clustocentric distance as an
environment indicator have found that the star-formation rates of
galaxies are correlated with the clustocentric distance in the following
ways: \textsl{(i)} the star-formation rates of galaxies near the centers 
of clusters are low and they rise with increasing clustocentric distance.
\textsl{(ii)} the correlation is steeper for galaxies with higher
star-formation rates, and \textsl{(iii)} there is a `characteristic
break' at a distance of 3-4 cluster virial radii at which the change in the
correlation is greatest \citep{lewis02a, gomez03a}.  It has been
suggested that gas-stripping plays a significant role in this rapid
truncation of star-formation, as well as the morphological
transformation that is observed in the cluster population
\citep{vogt04a}, although most statistical studies find that this
process cannot be extremely rapid \citep{kodama01a, balogh00a}.
Models generally find that the star-formation rates of cluster
galaxies depend primarily on the time elapsed since their accretion
onto the massive system, but that the cessation of star formation may
take place gradually over a few billion years \citep{balogh00a}.

In this short paper, we build upon previous studies in several ways.
We use the \citet{berlind06a} cluster catalog that was created with
a different algorithm than previous studies.  In particular, the algorithm 
did not use galaxy colors in the group and cluster identification,
so our galaxy colors are unbiased even in the cluster cores.  Moreover,
the cluster catalog was designed to recover clusters of galaxies that occupy
the same underlying dark matter halo.  Consequently, the abundance of
these clusters is unbiased relative to the abundance of halos and can
thus be used to obtain robust estimates of the cluster masses.  The method
we use to estimate cluster masses differs significantly from the methods
used in previous studies.  This difference leads to a different conclusion
regarding the activity observed in the theoretically active infall regions 
around clusters.  Finally, we examine the relationship between the 
morphology--environment relation and the color--environment relation.

We use the Sloan Digital Sky Survey (SDSS) data to measure the clustocentric 
distances of $\sim52,000$ galaxies relative to $\sim950$ clusters within the
redshift range $0.015<z<0.068$.  We use high quality photometric and 
spectroscopic data to examine the dependence of $^{0.1}[g-r]$ color, 
absolute magnitude $M_{^{0.1}i}$, \Halpha\ equivalent width (EW), 
concentration (S\'ersic index), and surface brightness $\mu_{^{0.1}i}$ 
on both clustocentric distance and cluster richness.  The large solid angular 
coverage of the SDSS permits the study of galaxies from cluster centers 
through the ``infall regions'' \citep{poggianti99a, balogh00a, kodama01a} 
and out to huge clustocentric distances.

In what follows, a cosmological world model with
$(\Omega_\mathrm{M},\Omega_\mathrm{\Lambda})=(0.3,0.7)$ is adopted,
and the Hubble constant is parameterized
$H_0\equiv100\,h~\mathrm{km\,s^{-1}\,Mpc^{-1}}$, for the purposes of
calculating distances and volumes with $h=1$ except where otherwise
noted \citep[\eg,][]{hogg99cosm}.

\section{Data}

The SDSS is taking $ugriz$ CCD imaging of $\sim10^4~\mathrm{deg^2}$ of
the Northern Galactic sky, and, from that imaging, selecting
$\sim10^6$ targets for spectroscopy, most of them galaxies with
$r<17.77~\mathrm{mag}$ \citep[\eg,][]{gunn98a,york00a,stoughton02a}.
All the data processing, including astrometry \citep{pier03a}, source
identification, deblending and photometry \citep{lupton01a},
calibration \citep{fukugita96a,smith02a}, spectroscopic target
selection \citep{eisenstein01a,strauss02a,richards02a}, spectroscopic
fiber placement \citep{blanton03a}, spectral data reduction and
analysis (Schlegel \& Burles, in preparation, Schlegel in preparation)
are performed with automated SDSS software.

Galaxy colors are computed in fixed bandpasses, using Galactic
extinction corrections \citep{schlegel98a} and $K$ corrections
\citep[computed with \texttt{kcorrect v3\_2};][]{blanton03b}.  They
are $K$ corrected, not to the redshift $z=0$ observed bandpasses, but to
bluer bandpasses $^{0.1}g$, $^{0.1}r$ and $^{0.1}i$ ``made'' by
shifting the SDSS $g$, $r$, and $i$ bandpasses to shorter wavelengths
by a factor of 1.1 \citep[\cf,][]{blanton03b, blanton03d}.  This means
that galaxies at redshift $z=0.1$ all have the same $K$ corrections:
$K(0.1)=-2.5 \log(1.1)$.

For the purposes of computing large-scale structure statistics, we
have assembled a complete subsample of SDSS galaxies known as the NYU
LSS \texttt{sample14}.  This subsample is described elsewhere
\citep{blanton05a}; it is selected to have a well-defined window
function and magnitude limit.  In addition, the sample of galaxies
used here was selected to have apparent magnitude in the range
$14.5<r<17.77~\mathrm{mag}$, redshift in the range $0.015<z<0.068$,
and absolute magnitude $M_{^{0.1}i}>-24~\mathrm{mag}$.  These cuts
left 52,569 galaxies.

A seeing-convolved S\'ersic model is fit to the azimuthally averaged 
radial profile of every galaxy in the observed-frame $i$ band, as
described elsewhere \citep{blanton03d,strateva01a}.  The S\'ersic
model has surface brightness $I$ related to angular radius $r$ by
$I\propto \exp[-(r/r_0)^{(1/n)}]$, so the parameter $n$ (S\'ersic
index) is a measure of radial concentration (seeing-corrected).  At
$n=1$ the profile is exponential, and at $n=4$ the profile is
de~Vaucouleurs.  In the fits shown here, values in the range
$0.5<n<5.5$ were allowed.

To every best-fit S\'ersic profile, the \citet{petrosian76a}
photometry technique is applied, with the same parameters as used in
the SDSS survey.  This supplies seeing-corrected Petrosian magnitudes
and radii.  A $K$-corrected surface-brightness $\mu_{^{0.1}i}$ in the
$^{0.1}i$ band is computed by dividing half the $K$-corrected
Petrosian light by the area of the Petrosian half-light circle.

The \Halpha\ line flux is measured in a 20~\AA\ width interval
centered on the line.  Before the flux is computed, a best-fit model
A+K spectrum \citep{quintero04a} is scaled to have the same flux
continuum as the data in the vicinity of the emission line and
subtracted to leave a continuum-subtracted line spectrum.  This method
fairly accurately models the \Halpha\ absorption trough in the
continuum, although in detail it leaves small negative residuals.  The
flux is converted to a rest-frame EW with a continuum found by taking
the inverse-variance-weighted average of two sections of the spectrum
about 150~\AA\ in size and on either side of the emission line.
Further details are presented elsewhere \citep{quintero04a}.

A caveat to this analysis is that the 3~arcsec diameter spectroscopic
fibers of the SDSS spectrographs do not obtain all of each galaxy's
light because at redshifts of $0.015<z<0.068$ they represent apertures
of between $0.6$ and $2.7~h^{-1}\,\mathrm{kpc}$ diameter.  The
integrity of our \Halpha\ EW measurement is a function of galaxy size,
inclination, and morphology.  We have looked at variations in our
\Halpha\ EW results as a function of redshift and found that the
quantitative results differ but our qualitative results remain the
same.

For each galaxy, a selection volume $V_\mathrm{max}$ is computed,
representing the total volume of the Universe (in
$h^{-3}~\mathrm{Mpc^3}$) in which the galaxy could have resided and
still made it into the sample.  The calculation of these volumes is
described elsewhere \citep{blanton03c, blanton03d}.  For each galaxy,
the quantity $1/V_\mathrm{max}$ is that galaxy's contribution to the
cosmic number density.

We use the group and cluster catalog described in \cite{berlind06a}.  
The catalog is obtained from a volume-limited sample of galaxies that is 
complete down to an $^{0.1}r$ band absolute magnitude of $M_r<-19$~mag 
and goes out to a redshift of 0.068.  Groups are identified using a
friends-of-friends algorithm (see e.g., Geller \& Huchra 1983; Davis
et al 1985) with perpendicular and line-of-sight linking lengths equal
to 0.14 and 0.75 times the mean inter-galaxy separation, respectively.
These parameters were chosen with the help of mock galaxy catalogs to
produce galaxy groups that most closely resemble galaxy systems that
occupy the same dark matter halos; \ie, that are bound.  The resulting
catalog contains 944 systems with a richness $\N \geq 5$ member
galaxies.  For consistency, we call these objects ``clusters''.  Note
that, unlike some other catalogs, galaxy colors are \emph{not} used in 
cluster identification.

\cite{berlind06a} calculate rough mass estimates for the clusters using 
the cluster luminosity function (where luminosity is defined as the
total luminosity in $M_r<-19$~mag galaxies in the cluster) and assuming 
a monotonic relation between a cluster's luminosity and the
mass of its underlying dark matter halo.  By matching the measured
space density of clusters to the theoretical space density of dark
matter halos (given the concordance cosmological model and a standard
halo mass function), they assign a virial halo mass to each cluster
luminosity.  The masses derived in this way ignore the scatter in mass
at fixed cluster luminosity and are only meant to be rough estimates.
The resulting masses for the clusters used here range from
$\sim10^{12}$ to $\sim10^{15}$ solar masses.  Each cluster has an
associated ``virial radius'' of
\begin{equation}
\Rvir =
\left(\frac{3}{4\,\pi}\frac{M}{200\,\rho_{o}}\right)^{\frac{1}{3}}
\quad ,
\end{equation}
where $M$ is the estimated mass of the cluster and $\rho_{o}$ is the
current mean density of the Universe.  Note that this method for
determination of the virial radii is very different from that employed
by other investigators.  Some have used a quasi-empirical formula
based on velocity dispersion \citep{gomez03a, christlein05a}, others
have assumed that cluster mass is directly proportional to richness
\citep{lewis02a}; in general these methods differ substantially, and
thus produce cluster catalogs with very different mass functions.

We use the cluster centers given by \citet{berlind06a}, which are
computed as the mean of the member galaxy positions.  We then calculate
the transverse projected clustocentric distance \Dcl\ from each galaxy 
to its nearest cluster center on the sky within 
$\pm1000~\mathrm{km\,s^{-1}}$ in radial velocity.
We split the cluster catalog into three richness bins.  The small,
medium, and large richness bins contain clusters with 5 to 9, 
10 to 19, and 20 or greater galaxies with $M_r<-19$~mag, respectively.  
We associate galaxies to their nearest large cluster, even if they are 
also nearby a smaller cluster; for example, a galaxy whose nearest 
cluster has a richness of $5\leq\N\leq 9$ falls into the ``small'' 
richness bin.  If this galaxy also has a nearby cluster of richness 
$10\leq\N\leq 19$ it would \emph{also} fall into the ``medium'' 
richness bin, but with a \Dcl\ which corresponds to this different cluster.

\section{Results}

\begin{figure*}
\begin{center}
\resizebox{!}{5.5in}{\includegraphics{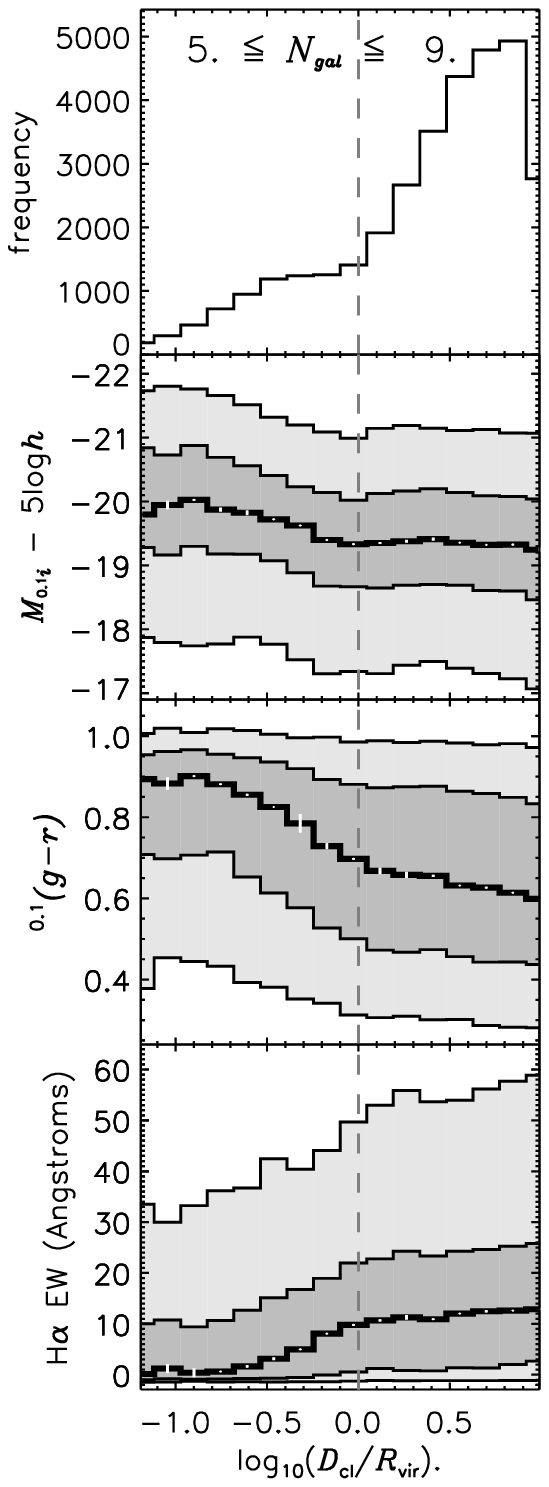}}%
\resizebox{!}{5.5in}{\includegraphics{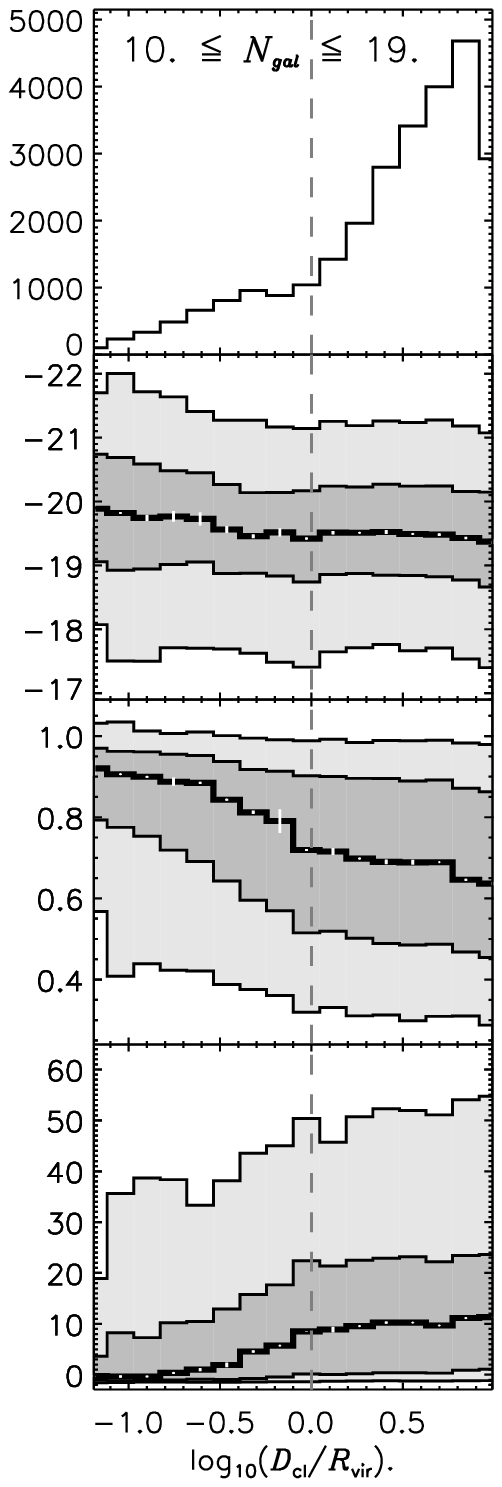}}%
\resizebox{!}{5.5in}{\includegraphics{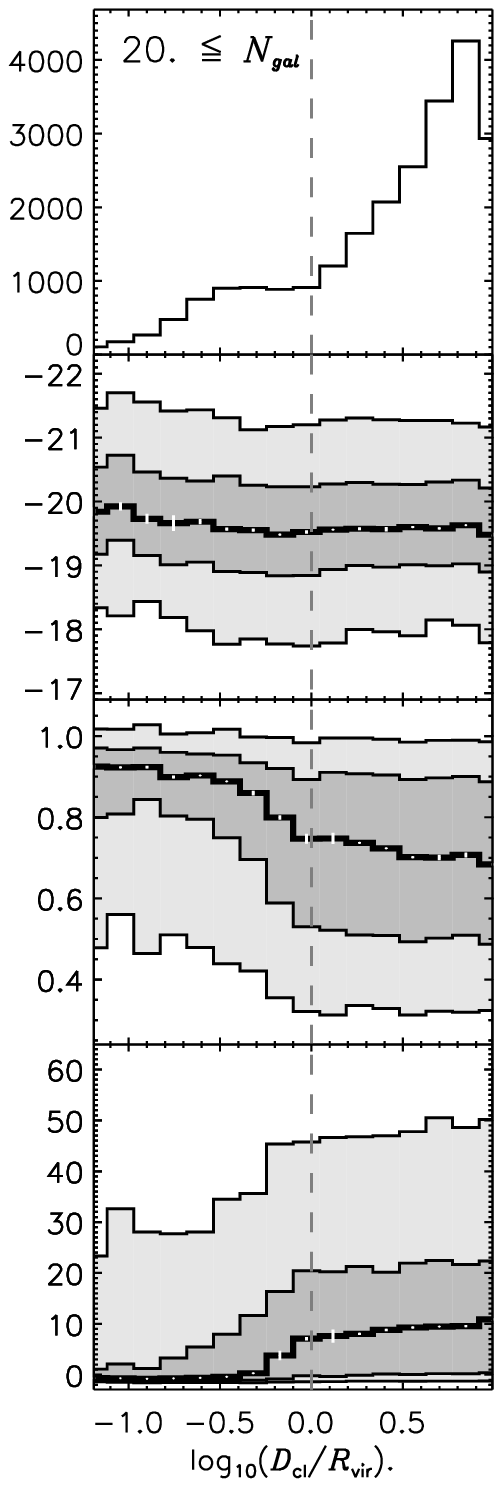}}
\end{center}
\caption{The dependence of the 5, 25, 50 (bold), 75, and 95 percent
$V_\mathrm{max}$-weighted quantiles of several galaxy properties on
\Dcl, shown for 3 different cluster-richness bins.  The ``weighted
quantile'' is defined as follows: The 5th percentile weighted quantile
has 5 percent of the total weight below it and 95 percent above it.
The top panel of each column shows the \Dcl\ distribution of galaxies
in the specified richness bin.  The other panels show the dependencies
of the quantiles of absolute magnitude $M_{^{0.1}i}$ , $^{0.1}[g-r]$
color, and \Halpha EW on \Dcl.  The figures are shaded to guide the
eye.  The errors in the 50 percent quantiles, calculated using the
jackknife method (10 jackknife trials, separated by constant
declination lines, in each of which 1/10 of the survey footprint is
dropped) are overplotted in white.
\label{fig:all_gals}}
\end{figure*}

We first examine the relation between three tracers of star-formation
history and clustocentric distance.  Figure~\ref{fig:all_gals} shows
the relationship between clustocentric distance and quantiles of three
$1/V_\mathrm{max}$-weighted galaxy properties: absolute magnitude
$M_{^{0.1}i}$, $^{0.1}[g-r]$ color and \Halpha\ EW for the three
different richness bins.  All three properties show a dependence on
clustocentric distance.  In agreement with previous work, we find that
galaxies near the centers of clusters are more luminous, redder and
lower in \Halpha\ EW than galaxies in the field.

For all three galaxy properties there is a change of slope or
``break'' in the quantile gradients at around one virial radius,
regardless of richness.  The break seen in \Halpha\ EW is similar to
that found by \cite{gomez03a}, except that they found this break at a
distance of 3-4 virial radii, whereas this figure shows it occurs at
$\Dcl\,\sim\,\Rvir$.  This discrepancy can be attributed to the
different methods of determining the virial radius.  Our estimates of
\Rvir\ are based on cluster abundances \citep{berlind06a}, whereas
\citet{gomez03a} use the \cite{girardi98a} approximation relating \Rvir\ 
to velocity dispersion.  We discuss this more in \S~4.

Although all these trends in galaxy properties display a break near
\Rvir, Figure~\ref{fig:all_gals} shows that the strength of the
trends depends on cluster richness, \N.  The different $M_{^{0.1}i}$
panels demonstrate that the $M_{^{0.1}i}$ \vs\ \Dcl\ gradient is
larger for the smaller clusters.  In other words, as one approaches
the center of a cluster, the increase in typical luminosity is greater
for smaller clusters.  The $^{0.1}[g-r]$ \vs\ \Dcl\ gradient depends
on cluster richness in the following two ways: \textsl{(i)}~the
transition in typical galaxy color from the cluster center (\ie,
redder) to the field value (\ie, bluer) is more abrupt and occurs
closer to \Rvir\ for larger clusters and \textsl{(ii)}~the color
distribution is overall narrower within larger clusters.  Similar to
color, the \Halpha\ EW \vs\ \Dcl\ gradient depends on cluster richness
in the following two ways: \textsl{(i)}~the transition in typical
\Halpha\ EW from the cluster center (\ie, lower) to the field value
(\ie, higher) is more abrupt and occurs closer to \Rvir\ for larger
clusters and \textsl{(ii)}~the \Halpha\ EW distribution is overall
narrower within larger clusters.

\begin{figure*}
\begin{center}
\resizebox{!}{5.5in}{\includegraphics{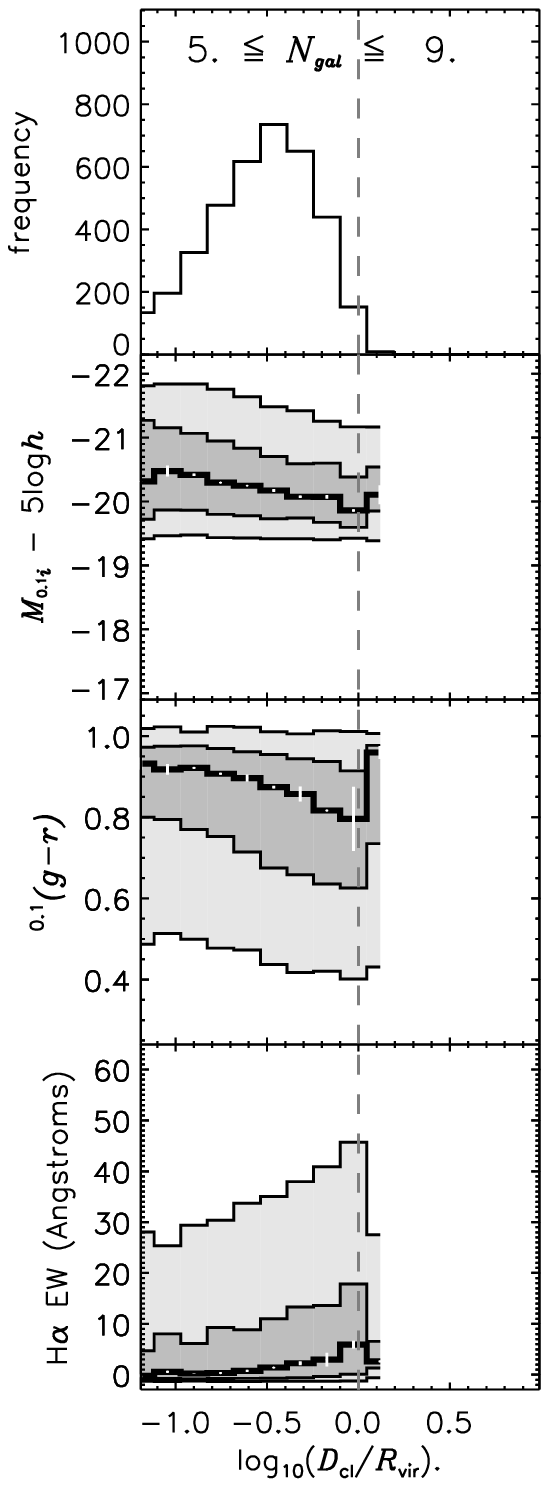}}%
\resizebox{!}{5.5in}{\includegraphics{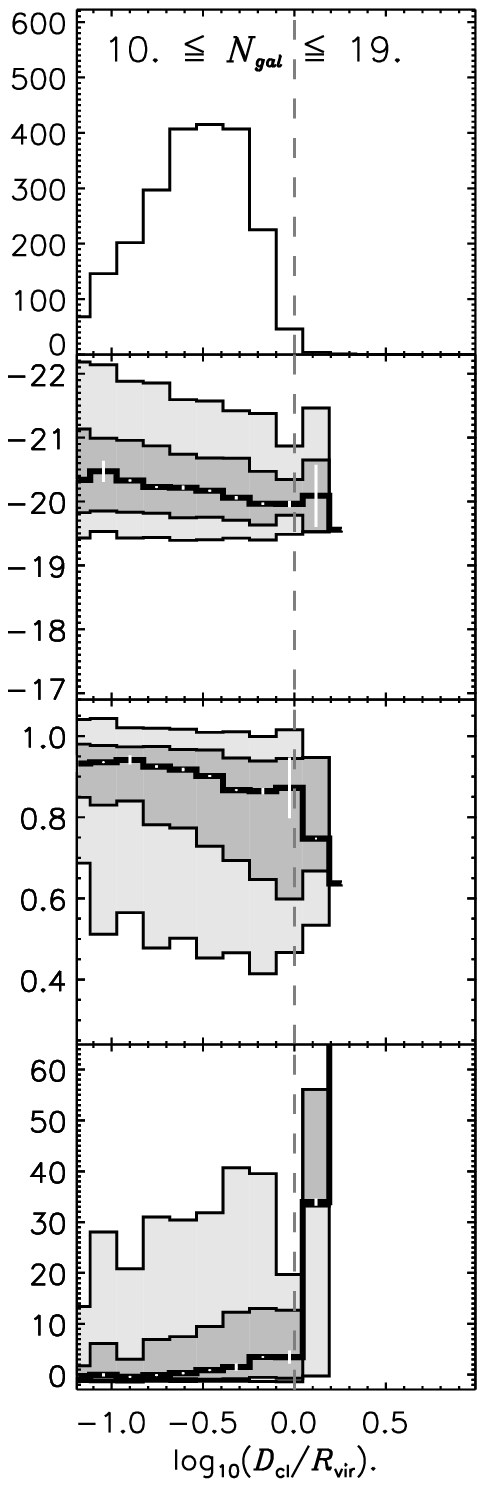}}%
\resizebox{!}{5.5in}{\includegraphics{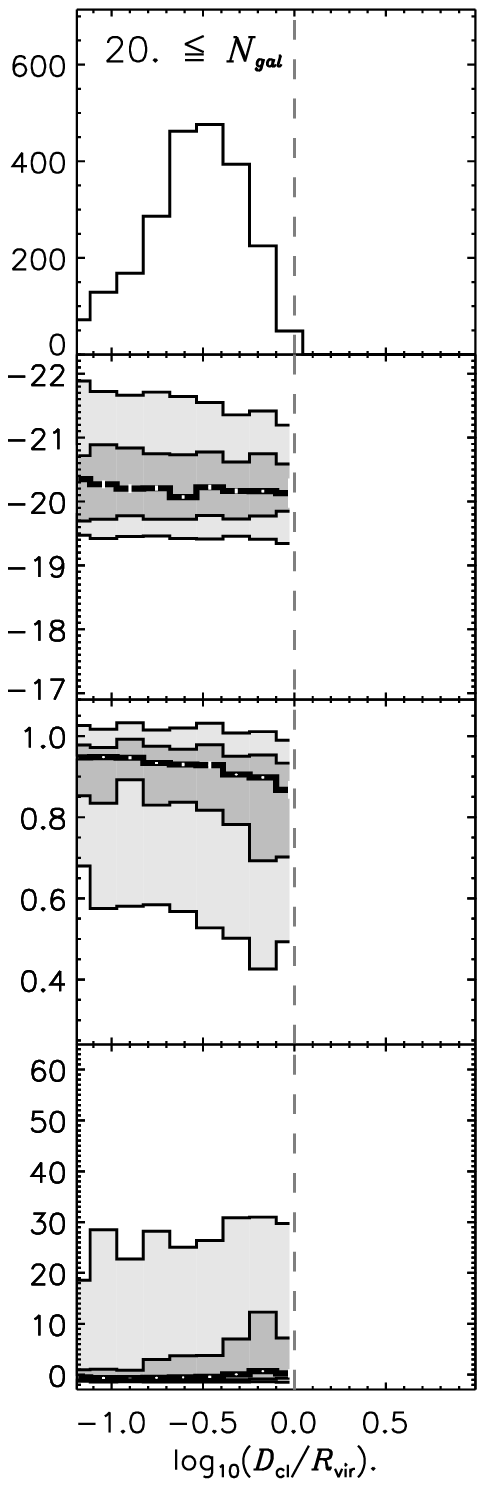}}
\end{center}
\caption{Similar to Figure~\ref{fig:all_gals}, except for galaxies that
\citet{berlind06a} have determined to be members of their closest clusters.
All trends are affected (relative to Figure~\ref{fig:all_gals}) by the
magnitude cut of $M_r<-19$~mag in the cluster definition; in
particular, the $M_{^{0.1}i}$ panels are strongly affected.
\label{fig:mem_gals}}
\end{figure*}

The plots in Figure~\ref{fig:all_gals} include all galaxies within
$\pm 1000~\mathrm{km\,s^{-1}}$ of each cluster center, so galaxies in
the foreground and background of the cluster contribute to this
Figure.  To investigate whether these projection effects (such as the
presence of ``infall interlopers'' discussed by \citealt{rines05a})
significantly influence the observed trends, we made the same plot in
Figure~\ref{fig:mem_gals}, but showing only the cluster members (\ie,
the galaxies that are members of their closest cluster).
Figure~\ref{fig:mem_gals} shows that there are still dependences on
clustocentric distance.  The trends are not as strong as in
Figure~\ref{fig:all_gals} because this subset of the data has a
smaller absolute magnitude range.  There is a cut of
$M_{^{0.1}r}<-19$~mag on this subset due to the cluster definition.
When we make the same $M_{^{0.1}r}$ cut on the sample used in
Figure~\ref{fig:all_gals}, the trends look the same as they do in this
Figure.  The result of this comparison suggests two things: first,
effects due to projection (\ie, due to galaxies with small \Dcl\ but
not spatially close to the cluster center) are small and don't affect
our results and, second, the fainter galaxies contribute more to these
trends than the brighter ones.

\begin{figure*}
\begin{center}
\resizebox{!}{5.5in}{\includegraphics{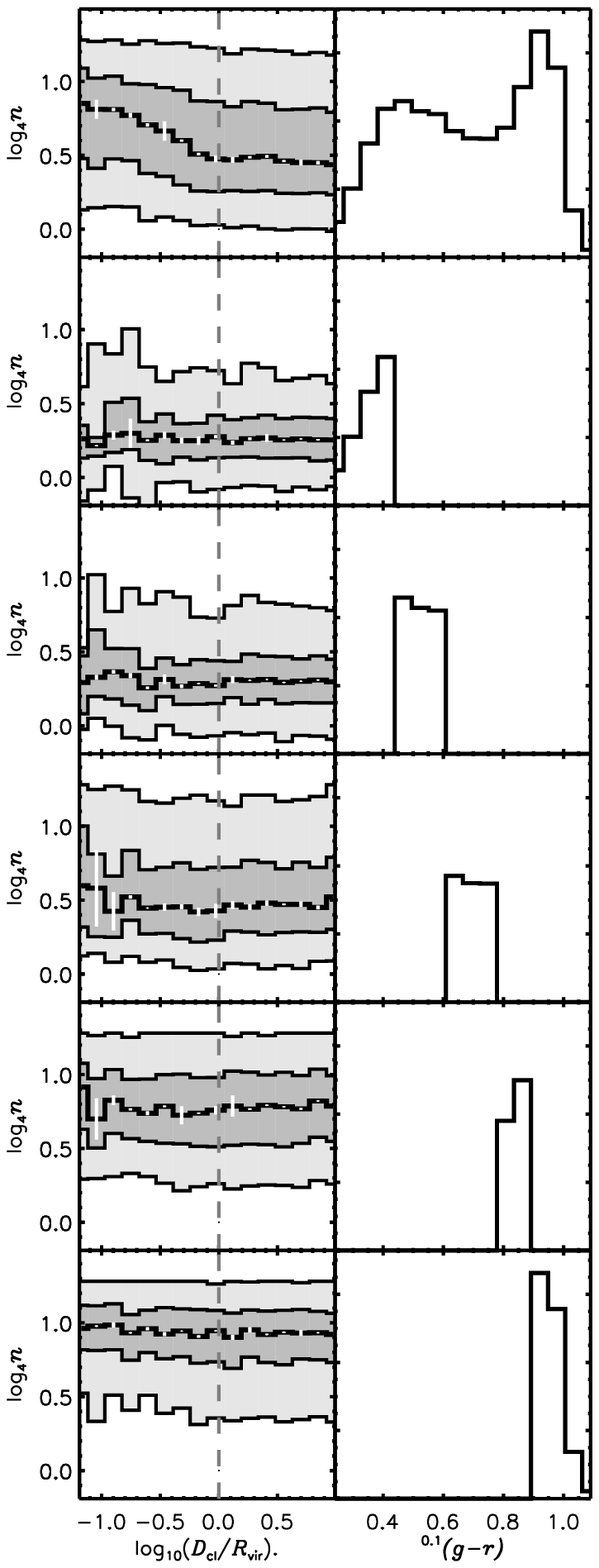}}%
\resizebox{!}{5.5in}{\includegraphics{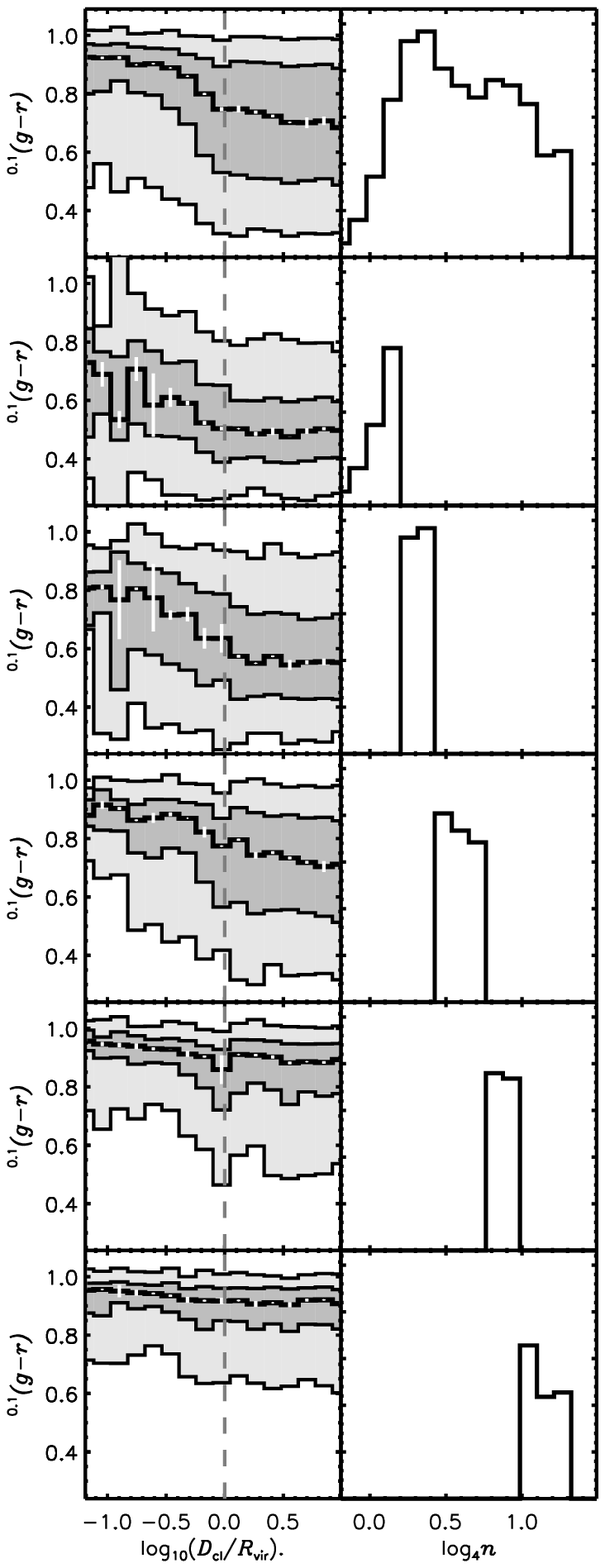}}
\end{center}
\caption{The dependence of radial concentration $n$ (left figure) and
$^{0.1}[g-r]$ color (right figure) on clustocentric distance \Dcl\ for
the smallest richness bin sample: $5\leq\N\leq 9$.  For the figure on
the left, the top-left panel shows how the quantiles of S\'ersic index
$n$ depend on \Dcl\ for the whole sample.  The top-right panel of this
figure shows the $^{0.1}[g-r]$ color distribution of the sample.  The
subsequent rows show the same for narrow color subsamples described by
the color distributions shown in the right column. All the panels in
the left column have the same layout as Figure~\ref{fig:all_gals}.
The figure on the right is very similar but with the $n$ and
$^{0.1}[g-r]$ properties interchanged.  Notice the dependence of
concentration on clustocentric distance almost vanishes in each color
subsample while the dependence of color on clustocentric distance
remains in each concentration subsample.  These phenomena were also
observed for the other two richness bin samples: $10\leq\N\leq 19$ and
$20\leq\N$.
\label{fig:n_dont_matter}}
\end{figure*}

Figures~\ref{fig:all_gals} and~\ref{fig:mem_gals} show how our tracers 
of star-formation history depend on clustocentric distance and cluster 
richness.  \cite{blanton05b} found that morphology tracers, such as
concentration and surface brightness, only depend on local overdensity
through their correlation with star-formation history.  This has not been 
demonstrated for clustocentric radius, which is a much higher signal-to-noise
environment estimator than local overdensity.  We investigate this in
Figure~\ref{fig:n_dont_matter}.

The left half of Figure~\ref{fig:n_dont_matter} shows the the relation 
between the S\'ersic index $n$ (concentration) and clustocentric 
distance for different color bubsamples (left panels), as well as the
1/Vmax-weighted distribution of $^{0.1}[g-r]$ color for each subsample
(right panels).  The dependence of concentration on clustocentric
distance for the whole sample is shown in the top row, and for narrow 
color subsamples in the following rows.  Similar to previous results
\citep{blanton05b}, we find that this dependence of concentration on
clustocentric distance almost completely vanishes within narrow color
subsamples.  The right half of Figure~\ref{fig:n_dont_matter} is similar, 
but with the concentration $n$ and $^{0.1}[g-r]$ color properties interchanged.
The dependence of color on clustocentric distance \emph{remains} even
within the narrow concentration subsamples.  We therefore conclude
that the dependence of concentration $n$ on clustocentric distance is
simply due to the dependence of concentration on color combined with the
dependence of color on clustocentric distance.  There is very little
\emph{independent} dependence of concentration on clustocentric
distance.  Surface brightness $\mu_{^{0.1}i}$ shows the same behavior
as concentration.

It is worth mentioning that galaxies that contain active galactic
nuclei (and therefore have \Halpha~EW not entirely due to star
formation) have a negligible effect on these results.  These results and
figures do not change substantially when we remove the AGN galaxies, as 
defined elsewhere \citep{kauffmann03b}.  We also note that ``edge effects''
from the survey boundaries do not alter our results; there is little
change to the figures and results when we remove galaxies and clusters
that lie near the survey edges.  Finally, our results are not sensitive
to the choice of definition for cluster centers.  We have verified that the
results do not change significantly if we assume that a cluster's center
is at the position of its most luminous galaxy, rather than the centroid of
all its galaxy positions.

\section{Discussion}

Using a complete sample of 52,569 galaxies in the redshift range of
$0.015<z<0.068$ we examine how $^{0.1}i$ band absolute magnitude
$M_{^{0.1}i}$, $^{0.1}[g-r]$ color, \Halpha\ EW, surface brightness,
and concentration are related to both the transverse projected 
clustocentric distance, \Dcl, and the richness, \N, of its nearest 
cluster.  Similar to previous results \citep{blanton05b,christlein05a}, 
we find that our morphology tracers (concentration and surface brightness) 
appear to be related to environment (\Dcl\, in this case) only indirectly 
through their relationships with the star-formation history tracers.  
This suggests that the well known morphology--environment relation is 
a residual of the star-formation-history--environment relation.  We note 
that, although our simple morphology indicators do not show direct dependence 
on clustocentric distance, this result does not disagree with that the
original morphology--density studies \citep{hubble36a, oemler74a,
dressler80a, postman84a} because these studies did not attempt to
separate morphological and star-formation dependences.  Our results
also do not disagree with previous work \citep{vogt04a} which suggests
that gas stripping plays a significant role in the morphological
transformation and rapid truncation of star formation by showing
asymmetries in HI and \Halpha\ flux on the leading edge of infalling
spiral galaxies.  There is no disagreement because our very blunt
morphology indicators are insensitive to these asymmetries.

We have shown that galaxies know about the distance to, \emph{and
richness of}, their nearest cluster and they express that knowledge
most clearly through their star-formation histories.  Some
morphological evidence suggests that they also know the direction as
well \citep{vogt04a}.

Where our results overlap those of previous investigators
\citep{lewis02a, gomez03a, blanton05b}, we mostly find good agreement.
In particular, we find that galaxy properties depend on clustocentric
distance much the same way as they do on other environmental
indicators \citep{blanton05b}.  Our result improves on this previous one
because the clustocentric distance is measured at much higher
signal-to-noise and probes environments on scales much smaller than
local overdensity measurements.  Where our results disagree with
previous work, it can mainly be attributed to the different methods of
computing cluster virial radii.  While it may seem that this is a
minor discrepancy, these disagreements become important at the
interpretation level.  In particular, the claim that clusters affect
galaxies well beyond the cluster virial radii \citep{lewis02a,
gomez03a} depends strongly on the calculation of virial radius.  Our
virial radii are estimated from cluster abundances (and an assumed 
``concordance'' cosmological model), whereas in previous work they
are estimated from cluster velocity dispersions.  In order to check the
difference between these two methods, we computed a second set of virial
radii for our clusters using the \cite{berlind06a} velocity dispersions
and the \cite{girardi98a} approximation that relates virial radius to
velocity dispersion.  We find that the virial radii based on velocity
dispersions are systematically lower than those based on abundances
by a factor of two to three.  This is sufficient to explain the difference
in interpretation between \cite{gomez03a} and this work.  We choose
to measure our virial radii using cluster abundances because the
\cite{berlind06a} cluster catalog was tuned to produce unbiased abundances
and we, therefore, believe them to be less subject to systematic errors 
than velocity dispersion estimates.  However, this point deserves to be 
studied in more detail.

After analyzing these results, we are in a position to address the
theoretically active ``infall region'' \citep{poggianti99a, balogh00a,
kodama01a} around virialized systems.  All three trends (color,
\Halpha, and absolute magnitude) as a function of \Dcl\ have a
``characteristic break'' at about the same virial-radius-normalized
distance from the cluster center but we find no clear evidence for an
``increase'' in star formation activity at the infall region.

Here we have investigated how galaxy colors, luminosities, and
star-formation rates (\Halpha\ EW) relate to clustocentric distance
and how these relations depend on cluster richness.  We find that for
larger clusters the transitions in the typical values of color and
star-formation rate from cluster centers to the field are more abrupt
and occur closer to the viral radius than those for smaller clusters.
We also find the increase in typical luminosity when looking from the
field to cluster centers is greater around smaller clusters.  A
question that naturally arises is: What physical physical processes
are involved in creating these richness-dependent variations?  It
could be that \textsl{(i)}~the possible transformation mechanisms are
stronger in larger clusters (\eg, more frequent tidal interactions or
a hotter intercluster medium), \textsl{(ii)}~galaxies in larger
clusters have been in this environment longer and therefore have had
more time to evolve to their long-term properties, or (more likely) a
combination of the two.

\acknowledgements We thank Alison Coil, Alister Graham, Erin Sheldon,
and Beth Willman for useful ideas, conversations, and comments on the
manuscript.  This research made use of the NASA Astrophysics Data
System.  ADQ, DWH, and MRB are partially supported by NASA (grant
NAG5-11669) and NSF (grant AST-0428465).

Funding for the creation and distribution of the SDSS Archive has been
provided by the Alfred P. Sloan Foundation, the Participating
Institutions, the National Aeronautics and Space Administration, the
National Science Foundation, the U.S. Department of Energy, the
Japanese Monbukagakusho, and the Max Planck Society. The SDSS Web site
is http://www.sdss.org/.

The SDSS is managed by the Astrophysical Research Consortium for the
Participating Institutions. The Participating Institutions are The
University of Chicago, Fermilab, the Institute for Advanced Study, the
Japan Participation Group, The Johns Hopkins University, Los Alamos
National Laboratory, the Max-Planck-Institute for Astronomy, the
Max-Planck-Institute for Astrophysics, New Mexico State University,
University of Pittsburgh, Princeton University, the United States
Naval Observatory, and the University of Washington.

\clearpage

\clearpage

\clearpage

\end{document}